\documentclass[journal]{IEEEtran}
\ifCLASSINFOpdf
\else
\fi

\usepackage{xcolor}
\usepackage{tabularx,booktabs}
\newcolumntype{C}{>{\centering\arraybackslash}X}
\newcommand{\hd}{\rotatebox{90}}
\usepackage{graphicx}

\newcommand{\numberOfStandardizedTests}{$12$} 
\newcommand{\numDepVars}{$19$}
\newcommand{\numIndepVars}{$927$}
\newcommand{\numParticipants}{$757$}

\newcommand{\blinded}[1]{\textsc{$($blinded-for-review$)$}}

\newcommand{\citeN}[1]{\cite{#1}}

\newcommand{\VarIRB}{IRB}
\newcommand{\VarITP}{ITP}
\newcommand{\VarOCB}{OCB}
\newcommand{\VarDevInter}{Interpersonal Deviance}
\newcommand{\VarDevOrg}{Organizational Deviance}
\newcommand{\VarAbstraction}{Abstraction}
\newcommand{\VarVocabulary}{Vocabulary}
\newcommand{\VarPerExtraversion}{Extraversion}
\newcommand{\VarPerAgreeableness}{Agreeableness}
\newcommand{\VarPerConscientiousness}{Conscientiousness}
\newcommand{\VarPerNeuroticism}{Neuroticism}
\newcommand{\VarPerOpenness}{Openness}
\newcommand{\VarAffectPos}{Positive Affect}
\newcommand{\VarAffectNeg}{Negative Affect}
\newcommand{\VarAnxiety}{Anxiety}
\newcommand{\VarAlcohol}{Alcohol}
\newcommand{\VarTobacco}{Tobacco}
\newcommand{\VarPhysical}{Physical Activity}
\newcommand{\VarSleep}{Sleep}

\newcommand{\diagonaltext}[1]{\rotatebox[origin=c]{90}{#1}}

\usepackage{algorithm,algorithmic}

\usepackage{multirow}

\usepackage{float}
\floatstyle{plaintop}
\restylefloat{table}

\usepackage{lscape}
\usepackage{cite}

\makeatletter
\def\bstctlcite{\@ifnextchar[{\@bstctlcite}{\@bstctlcite[@auxout]}}
\def\@bstctlcite[#1]#2{\@bsphack
 \@for\@citeb:=#2\do{%
    \edef\@citeb{\expandafter\@firstofone\@citeb}%
    \if@filesw\immediate\write\csname #1\endcsname{\string\citation{\@citeb}}\fi}%
 \@esphack}
\makeatother

\begin{document}
%

\title{Jointly 
Predicting Job Performance, Personality, Cognitive Ability, Affect, and 
Well-Being \thanks{\copyright~2020 IEEE. Personal use of this material is permitted. Permission from IEEE must be obtained for all other uses, in any current or future media, including reprinting/republishing this material for advertising or promotional purposes, creating new collective works, for resale or redistribution to servers or lists, or reuse of any copyrighted component of this work in other works}
}
%
%
%


\author{\IEEEauthorblockN{
Pablo Robles-Granda\IEEEauthorrefmark{1}, \and 
Suwen Lin\IEEEauthorrefmark{1}, 
\and
Xian Wu\IEEEauthorrefmark{1}, \and
Sidney D'Mello\IEEEauthorrefmark{2}, \and
Gonzalo J. Martinez \IEEEauthorrefmark{1}, \and
Koustuv Saha\IEEEauthorrefmark{5}, \and
Kari Nies\IEEEauthorrefmark{3}, \and
Gloria Mark\IEEEauthorrefmark{3}, \and
Andrew T. Campbell\IEEEauthorrefmark{4}, \and
Munmun De Choudhury\IEEEauthorrefmark{5}, \and
Anind D. Dey\IEEEauthorrefmark{6}, \and
Julie Gregg\IEEEauthorrefmark{2}, \and
Ted Grover\IEEEauthorrefmark{3}, \and
Stephen M. Mattingly \IEEEauthorrefmark{1}, \and
Shayan Mirjafari\IEEEauthorrefmark{3}, \and
Edward Moskal\IEEEauthorrefmark{1}, \and
Aaron Striegel\IEEEauthorrefmark{1}, \and
Nitesh V. Chawla\IEEEauthorrefmark{1}\IEEEauthorrefmark{8} \\
 \IEEEauthorblockA{\IEEEauthorrefmark{1}University of Notre Dame
 }
 \IEEEauthorblockA{\IEEEauthorrefmark{2}University of Colorado, Boulder 
}
 \IEEEauthorblockA{\IEEEauthorrefmark{3}University of California, Irvine 
}
 \IEEEauthorblockA{\IEEEauthorrefmark{4}Dartmouth College 
}
 \IEEEauthorblockA{\IEEEauthorrefmark{5}Georgia Institute of Technology 
}
 \IEEEauthorblockA{\IEEEauthorrefmark{6}University of Washington 
}\\
 \IEEEauthorblockA{\IEEEauthorrefmark{8}Corresponding Author: {\tt{nchawla@nd.edu}}
}}}

%
%

\markboth{}{Robles-Granda \MakeLowercase{\textit{et al.}}: Jointly Predicting of Health, Job Performance, and Psychometric Information and Wellness}
%



\maketitle

\begin{abstract}

Assessment of job performance, personalized health and psychometric measures are domains where data-driven and ubiquitous computing exhibits the potential of a profound impact in the near future. Existing techniques use data extracted from questionnaires, sensors (wearable, computer, etc.), or other traits, to assess well-being and cognitive attributes of individuals. However, these techniques can neither predict individual’s well-being and psychological traits in a global manner nor consider the challenges associated to processing the data available, that is incomplete and noisy. 
In this paper, we 
create a benchmark for predictive analysis of individuals from a perspective that integrates: 
physical and physiological behavior, psychological states and traits, and job performance.
We design data mining techniques as benchmark
and uses real noisy and incomplete data derived from wearable sensors to predict 19 constructs 
based on \numberOfStandardizedTests~standardized well-validated tests.
The study included \numParticipants~participants who were knowledge workers in organizations across the USA with varied work roles. 
We developed a data mining framework to extract the meaningful predictors for each of the \numDepVars~variables under consideration. 
Our model is the first benchmark that combines these various instrument-derived variables in a single framework to understand people's behavior
by leveraging real uncurated data from wearable, mobile, and  social media sources. 
We verify our approach experimentally using the data obtained from our longitudinal study. The results show that our framework is consistently reliable and capable of predicting the variables under study better than the baselines when prediction is restricted to the noisy/incomplete data.
\end{abstract}


\section{Introduction}
\bstctlcite{IEEEexample:BSTcontrol}

Understanding health and well-being will be fundamental for the workplace of the future. With the advent of wearable devices that can track individual in situ patterns of activity and affect, precise and objective measures of health and well-being can be obtained. The collection of such precision data could benefit individuals and also organizational efforts to develop programs to promote well-being of employees, which can provide economic advantages for organizations \cite{abbas2016personalized,goetzel2014workplace,aldana2005financial}. Such detailed and continual collection of data can yield valuable insights into health and wellness behavior, e.g. on employee stress, on how sleep patterns impact work performance, and physical activity\cite{yang2009relationship,sano2017designing,smyth2018everyday,quante2019seasonal}. Further, such data has utility and reusability for future studies \cite{Goldacre2016}.

In recent years there has been an expanding body of literature that assesses well-being and its effect on productivity. On the one hand, personalized well-being assessment \cite{Chawla2013,abbas2016personalized,andreu2015wearable}\footnote{an expanded list of related works is presented in Section \ref{sec:related-work}} is receiving more attention due to the availability of sensor data. On the other hand, it is known that well-being, in conjunction with personality, cognitive, and personality traits, can affect job performance \cite{wright2000psychological,chiaburu2011five,kamdar2007joint,barrick1991big,barrick2001personality,eysenck2012model,salgado1997five,tett1991personality}, which in turn can affect organizations more broadly. 
Sensor data could play a major role in the latter, as it could be applied to current efforts of automating the evaluation of jobs, skills, and wages \cite{NSF,McKinsey,schneider2018empowerment}.

Despite advances that have been achieved in the design of computational methods both to collect and analyze wearable data to assess 
well-being, e.g.,\cite{schaule2018employing,olguin2009capturing,Chawla2013,abbas2016personalized,andreu2015wearable}, substantial challenges still remain. These 
studies have relied on data gathered under specific conditions using assumptions such as: small samples, certain populations (homogeneous demographics and work roles), or  controlled environments (within particular scenarios and locations). Furthermore, when associated with prediction of social constructs, such as productivity or performance, modeling well-being becomes more complex and difficult to achieve even when long-term longitudinal psychometric data is used \cite{Salganik201915006}. One reason is because it depends on situational, contextual, and personal information ($\!\!$\cite{judge2015person,viswesvaran2000perspectives}) that is not always readily available due to privacy and other constraints,  \cite{wright2000psychological,chiaburu2011five,kamdar2007joint,barrick1991big,barrick2001personality,eysenck2012model,salgado1997five,tett1991personality}. Overcoming these issues and maximizing the utility of multi-sensor data to assess well-being and workplace performance will require generalizable strategies built for diverse populations, continual data collection in a wide range of work settings, with larger samples of heterogeneous individuals, from various geographic locations and job environments, and from multimodal sources that provide a more thorough view of physical and behavioral patterns in an unobtrusive manner. Predicting well-being and workplace performance and taking advantage of this multimodal sensor data requires models that are robust to the messiness of real-world data due to missing or noisy entries \cite{mallinckrodt2003assessing,molenberghs2004analyzing}.

Our Tesserae Project \cite{mattingly2019tesserae} is the first to create a sensing system that fuses a comprehensive suite of broad modalities for automated modeling of individual physical and psychological differences as well as job performance. Figure \ref{fig:sensors} shows a diagram of the personal, social, contextual, and specialized sensors we used to gather data in an unobtrusive manner for this project. These sensors collect information that is representative of an individual's behaviors, physical attributes, and mental states, in addition to the context and interactions related to job and daily activities, which include offline and online interactions, phone and computer usage, and social media use. In order to obtain rich data for Tesserae, we consider a diverse cohort of \numParticipants~individuals, all knowledge workers across the USA from various organizations and with varied work roles, in both home and work environments. 

 \emph{Research Questions.} In this paper, we consider three core issues in order to create a truly global predictive model of well-being and psychological traits: 1) How do we integrate data from different sensors and modalities?  2) How do we develop machine learning methods that can deal with the challenges of multi-modality, varying levels of missingness and noise, and inter-individual and inter-sensor variance?  3) Leveraging these data and models, can we discover and predict traits about individual job performance, well-being, and personality? 
  
  \emph{Data Challenge.} The first challenge these research questions pose is the nature of the data. Real-world sensor and wearable data is messy due to factors such as: missingness, varying compliance rates from the participants in the study, data originating from multimodal interaction, different individual baselines and (ir)regularities, temporal variations, and generalizability issues due to the fact that, while models are build on a subset of the data, the evaluation and application of the predictions may be done on a truly blinded testing set that may not be representative of the training data. While sensor data have been used to predict and assess human behaviour and well-being \cite{madan2010social,yuan2013we,calabrese2013understanding,Mariani:2013}, these predictions are done on highly curated and homogeneous data. Thus, such models may be over-optimistic about what can be achieved in real scenarios.

\emph{Modeling Challenge.} We address these data issues while also bringing to the fore challenges for machine learning algorithms with the task of predicting a variety of job performance, psychological and well-being outcomes. 
Specifically, we model \numDepVars~constructs that can be categorized into three groups: physical/physiological, psychological, and job performance constructs \cite{williams1991job,griffin2007new,fox2012deviant,bennett2000development,Shipley2009,soto2017next,watson1999panas,Spielberger:83,saunders1993development,palipudi2016methodology,craig2003international,buysse1989pittsburgh}.  
The \numIndepVars~predictors were obtained from sensors assigned to each participant: a wearable (Garmin VivoSmart 3), a phone agent (an app for iPhone and Android), 4 beacons (office, home, and two portable), and social media data (Facebook). These predictors are filtered out with dimensionality reduction and feature selection techniques to extract subsets of meaningful predictors for each of the \numDepVars~variables under consideration.

As described before (and further detailed in Section \ref{sec:related-work}), sensor data have been used to create models of human behaviour, including: daily activities, mobility, \cite{madan2010social,yuan2013we,calabrese2013understanding}, well-being  \cite{Mariani:2013}, and academic performance \cite{schaule2018employing}. However, to our knowledge there is no study that combines various data sources in a single comprehensive analysis, while addressing several data messiness and algorithmic challenges to achieve the goal of creating a single framework to predict people's behavior, physical and psychological well-being, and job performance.

To achieve our complex goal, we combine four main ideas. 
First, we consider various imputation approaches as well as co-dependencies among the variables for both future selection and to predict other constructs (question 1) that, unlike other approaches that deal with missing data for longitudinal scenarios \cite{mallinckrodt2003assessing}, do not require complex likelihood-based techniques. Second, we use a fusion technique to synthesize the multi-modal sensor-derived features, i.e., predictor variables (question 1). Third, we consider an ensemble learning technique that incorporates various machine learning models to provide a comprehensive view of the participants' behaviors (question 2). Fourth, in addition to our data policy and machine learning design 
we use higher order networks (HON) \cite{Xue1600028} to obtain descriptions of each individual and how they relate to job performance, well-being, and personality (question 3).

We evaluate our approach experimentally to verify if it is possible to predict the variables that encapsulate our global description of individuals. To ensure generality of our results we perform 5-fold cross validations at all stages of the model construction: feature selection, dimensionality reduction, tests, blinded-validations, and reliability analysis. 

In summary, our contributions are the following: 1) We provide a framework whereby noisy, heterogeneous, multimodal data can be fused without the need for highly specialized and unrealistic curation or trimming of the data; 2) We provide a benchmark that leverages the data fused from the various modalities to produce more integrated predictions of human behavior than existing techniques; and 3) We implement our benchmark and verified experimentally its capability to predict job performance, well-being, and personality by testing both how accurate and reliable our benchmark is when applied to the data obtained from our longitudinal study. The results show that our sensor-based framework and benchmark perform favorably with respect to theory-driven (survey based) baselines. We verify the results using various reliability tests.

\section{Background}
\label{sec:background}

 In order to acquire a comprehensive view of an individual physical attributes, psychological features, personality, and job performance, we used a set of psychometric surveys. 
 We administered this battery of surveys at the beginning of the study, and periodically afterwards over the first 60 days of this year-long study. The variables that we predict were extracted from these surveys is listed in Table \ref{tab:general-gt-variables} and are the following (grouped per variable type):
 
\subsection{Job Performance}

We considered five variables that assess job performance from three perspectives: task performance, organizational citizenship behavior, and counterproductive work behavior 
\cite{viswesvaran2000perspectives,rotundo2002relative,borman1993expanding,cortina2012personnel,dalal:09}.
The surveys we use are based on studies about behaviors that lead to achieving organizational goals as in \cite{motowidlo2012job,campbell1990modeling,campbell2015modeling}.

\subsubsection{Task Performance}
We measure task performance using two variables: 
In-Role Behavior (IRB) \cite{williams1991job} and Individual Task Proficiency (ITP) \cite{griffin2007new}.
The former measures an individual's perception of her job performance based on completion of tasks associated to the position of the individual in the organization. The latter measures the individual's perception of how frequently she completed the core tasks of her job in the last month, how frequently these were completed well, and how frequently she verified these were completed well. 
Both IRB and ITP are instruments validated with significant samples \cite{williams1991job,griffin2007new}.

\subsubsection{Organization Citizenship Behavior}
Organizational citizenship behavior was assessed using the Organizational Citizenship Behavior Checklist (OCB-C;\cite{fox2012deviant}). OCBs are optional actions that are not rewarded by a worker's organization.
OCB-C is also a validated instrument as described in \cite{fox2012deviant}.

\subsubsection{Counterproductive Work Behavior}
(CWBs) are actions that purposefully harm either the organization or individuals within the organization \cite{sackett2002structure}. To measure CWB we use the Interpersonal and Organizational Deviance (IOD) scale \cite{bennett2000development}. The instrument is broken into two major categories of items: (1) Interpersonal Deviance and (2) Organizational Deviance. The IOD is a $19$ item survey, $7$ for Interpersonal Deviance and $12$ for Organizational Deviance. Each item has a seven-point frequency score: 1 - never, 7 daily. In our predictions we consider each major category as a separate variable.
IOD was validated as described in \cite{bennett2000development}.

\subsection{Psychological Constructs}

We focus on four psychological constructs: cognitive ability, personality, affect, and anxiety. 

\subsubsection{Cognitive Ability}

We consider the Shipley Institute of Living Scales 2 (Abstraction and Vocabulary sub-tests) \cite{Shipley2009}. We use this 
test to 
measure the fluid and crystallized intelligence respectively \cite{cattell1987intelligence,schneider2012cattell}. See \cite{schmidt2004general} for a study on the relation of cognitive ability and job performance.
Shipley 2 has high reliability and internal consistency \cite{Shipley2009}.

\subsubsection{Personality}

Personality was measured in the initial ground truth battery via the Big Five Inventory-2 (BFI-2; \citeN{soto2017next}). 
The Big Five Personality Traits and their main characteristics are:
Extraversion,
Agreeableness,
Conscientiousness,
Neuroticism, and  
Open-Mindedness.
Each of the Big Five Personality Traits have varying levels of association to job performance \cite{anderson1998update,barrick1991big,barrick2001personality,salgado1997five,tett1991personality,bartram2005great,feist1998meta}.
BFI-2 was validated with four datasets \cite{soto2017next}.


\subsubsection{Affect}

We use the Positive and Negative Affect Schedule-Expanded Form (PANAS-X; \citeN{watson1999panas}). Affect variation is a key indicator of a person's mental health and is critical for job performance and other job behaviors \cite{chen1992relationships,penney2005job,fox2001counterproductive,fox1999model}.
\cite{mackinnon1999short} showed that PANAS-X is a reliable instrument.

\subsubsection{Anxiety}

We use the State-Trait Anxiety Inventory (STAI; \citeN{Spielberger:83}). Anxiety is another key indicator of a person's mental health. 
STAI was validated by \cite{Spielberger:83}.

\subsection{Health and Physical Variables}

\subsubsection{Alcohol Consumption}

We use the Alcohol Use Disorders Identification Test (AUDIT), which was developed by the World Health Organization (WHO; \citeN{saunders1993development}). The effects of alcohol consumption on job performance and other areas of people's lives are well documented \cite{mullahy1992effects,podolsky1985investigating,giancola1998executive,galanter1998recent,hussong2001specifying,paunonen2003big}.
AUDIT validity has been widely studied for instance in \cite{bohn1995alcohol,piccinelli1997efficacy,volk1997alcohol}.

\subsubsection{Physical Activity}

We use the International Physical Activity Questionnaire (IPAQ;  \citeN{craig2003international}). Physical activity affects not only physical health but also mental wellness and job performance \cite{pronk2004association,ratey2011positive,toker2012job,salmon2001effects}.
The test-retest reliability for the IPAQ questionnaires is reported in \cite{craig2003international}.

\subsubsection{Sleep}

We use the Pittsburgh Sleep Quality Index (PSQI; \citeN{buysse1989pittsburgh}). It is critical to incorporate sleep 
because poor sleep directly impacts job performance \cite{rosekind2010cost}, 
 cognitive ability, and mental health \cite{barber2013better,barnes2012working}.
 PSQI has both good internal reliability and good test-retest reliability \cite{buysse1989pittsburgh}.

\subsubsection{Tobacco Use}
We used a modified version of the Global Adult Tobacco Survey GATS from the World Health Organization (WHO, \citeN{palipudi2016methodology}). Tobacco use is associated with stress, negative emotionality, lower agreeableness, 
\cite{parrott1999does,paunonen2003big,mccrae1978anxiety,cherry1976personality,kassel2003smoking,ikard1969scale}, and work performance \cite{halpern2001impact}. We use a modified version of GATS with three items: whether the participant is a current smoker, if they use tobacco daily, and the quantity used in the past week. We predict the last item only.
GATS was reviewed and approved by the GATS Questionnaire Review Committee of the WHO.

\begin{table}[]
\centering
\begin{tabular}{lll}
\hline
Type                             & Subtype                                   & Variable             \\ \hline
\multirow{5}{*}{\diagonaltext{$\begin{array}{c}
\mbox{Job}\\
\mbox{Performance}\\
\end{array}$}} & \multicolumn{1}{c}{\multirow{2}{*}{Task}} & IRB\cite{williams1991job}                   \\ \cline{3-3} 
                                 & \multicolumn{1}{c}{}                      & ITP\cite{griffin2007new}                   \\ \cline{2-3} 
                                 & Org. Cit. Behavior             & OCB\cite{fox2012deviant}                   \\ \cline{2-3} 
                                 & \multirow{2}{*}{Deviance \cite{bennett2000development}}                & Interpersonal                   \\ 
                    \cline{3-3} & &  Organizational                  \\
                    \hline
\multirow{10}{*}{ \diagonaltext{Psychological}}  & \multirow{2}{*}{Cognitive 
\cite{Shipley2009}}        & Vocabulary    \\ \cline{3-3} 
                                 &                                           & Abstraction   \\ \cline{2-3} 
                                 & \multirow{5}{*}{Personality \citeN{soto2017next}}              & Extraversion          \\ \cline{3-3} 
                                 &                                           & Agreeableness         \\ \cline{3-3} 
                                 &                                           & Conscientiousness     \\ \cline{3-3} 
                                 &                                           & Neuroticism \\ \cline{3-3} 
                                 &                                           & Openness       \\ \cline{2-3} 
                                 & \multirow{2}{*}{Affect\citeN{watson1999panas}}                                    & Positive              \\ \cline{3-3} 
                                 &                                           & Negative              \\ \cline{2-3} 
                                 & Trait Anxiety \citeN{Spielberger:83}                                    & Anxiety                 \\ \cline{1-2} \cline{3-3} 
\multirow{4}{*}{\diagonaltext{Health}}          & \multirow{2}{*}{Consumption}                       & \VarAlcohol\citeN{saunders1993development}                 \\ \cline{3-3} 
                                 &                               & \VarTobacco\citeN{palipudi2016methodology}         \\
                                 \cline{2-3} 
                                 & \multirow{2}{*}{ Activity}                         & Physical\citeN{craig2003international}                  \\ \cline{3-3} 
                                 &                                     & \VarSleep\citeN{buysse1989pittsburgh}                  \\ 
\hline
\end{tabular}
    \caption{List of Dependent Variables}
    \label{tab:general-gt-variables}
\end{table}

\begin{table}[h!]
    \centering
    \begin{tabular}{l|r|r }
Variable Group  & {Sensor/Social}  & Attributes \& Traits\\
\hline
Job Performance & \cite{schaule2018employing}\cite{olguin2009capturing} \cite{mirjafari2019differentiating}
&     \cite{barrick1991big,eysenck2012model,salgado1997five,higgins2007prefrontal,barrick2001personality,tett1991personality,ng2008relationship,chiaburu2011five,kamdar2007joint,schmidt2014general} 
\\
\hline
Cognitive Ability & \cite{tamez2019ensemble,alloway2012impact}  &     \cite{moutafi2005can} \\
Personality &	\cite{bai2012big,skowron2016fusing,shen2013understanding,abadi2015inference,olguin2009capturing}  &     \cite{tetlock1987accountability,schmidt2014general}  \\
Affect &	\cite{healey2000wearable,ghandeharioun2017objective,liu2008online,tuarob2017you}   &     \cite{tuarob2017you,fayard2012exploring} \\
\VarAnxiety &	\cite{zheng2016unobtrusive,ghandeharioun2017objective,liu2018autonomic}   &    \cite{dalal2014within,spielberger2009assessment,Spielberger:83}
\\
\hline
\VarAlcohol &	\cite{mumtaz2016automatic,marques2009field}   &    
\cite{Lindgren:2019}  \\
\VarTobacco &	\cite{hovell2019randomised}   &    \cite{ridner2005predicting}  \\
\VarPhysical &	\cite{choudhury2008mobile,plasqui2007physical}   &    \cite{vDyvk2015} \\
\VarSleep &	\cite{min2014toss,sano2015prediction,razjouyan2017improving,staples2017comparison}   &    \cite{staples2017comparison}
\\
    \end{tabular}\\
    \caption{Literature per Variable Group and Predictor Types}
    \label{tab:related-work}
\end{table}

\section{Related Work}
\label{sec:related-work}

We divide the related work into sections associated with each of the categories of dependent variables.

\emph{Psychological Variables.}
The work 
of \cite{moutafi2005can} that estimates cognitive ability based on other features such as personality traits.  \cite{alloway2012impact} showed that social media engagement was predictive of performance on some cognitive attributes such as working memory, attentional control, and others. \cite{tamez2019ensemble} used MRI to predict fluid intelligence using MRI data. 
According to \cite{schmidt2014general} personality (introversion) and abstraction are the cause for intellectual curiosity and intellectual curiosity, in addition to other traits, is the cause for vocabulary (crystallized intelligence). 
Personality is estimated by \cite{bai2012big,skowron2016fusing,shen2013understanding} using the individual's interactions with computers. However, it is more common to use other individuals traits to predict personality, for instance accountability \cite{tetlock1987accountability}. Affect is estimated in the work of \cite{healey2000wearable,ghandeharioun2017objective,abadi2015inference,liu2008online}. 
\cite{tuarob2017you} predicted positive and negative affect using wearables, in addition to perception of health and satisfaction with health and life. 
\cite{fayard2012exploring} showed that there is a relation between possitive affect and conscientiousness. 
Anxiety is predicted by \cite{zheng2016unobtrusive,ghandeharioun2017objective}. 
Techniques that predict stress and anxiety based on ECG monitoring have also been proposed 
\cite{liu2018autonomic}. 
\cite{schmidt2014general} also found that crystallized intelligence is the main cause for mental health at later stages in life.

\emph{Physical Variables and well-being} assessments are related to habits of participants. For instance, alcohol consumption is predicted by \cite{mumtaz2016automatic,marques2009field}. 
using 
EEG signals and trans-dermal devices. 
Alcohol consumption is related to self-control and working memory capacity \cite{Lindgren:2019}. 
Physical activity (IPAQ) is predicted by \cite{choudhury2008mobile} using mobile sensing. Other alternatives include more specialized devices such as accelerometers \cite{plasqui2007physical}. 
\cite{vDyvk2015} show that physical activity patterns earlier in life can predict some activity patterns later in life.
Sleep quality is estimated using wearables by \cite{min2014toss,sano2015prediction}. There are also more invasive techniques such as the one in \cite{razjouyan2017improving}. More specialized work to estimate sleep for patients with schizophrenia also exists (see a comparative analysis in \cite{staples2017comparison}). To our knowledge, tobacco consumption has been monitored using air sensors as in \cite{hovell2019randomised} but not wearable sensors. Also, \cite{ridner2005predicting} can predict the smoker group membership (never, established, former, non-daily, and daily) using family history, depression, consumption of other substances, and demographics. In general, all the physical well-being assessment techniques were developed in isolation.

\emph{Job Performance}
itself is usually measured through either subjective rating scales \cite{campbell1990modeling,sonnentag2008job} or objective performance outcomes, such as sales amounts, production numbers, etc. \cite{campbell1990modeling}. Using wearable sensor data to estimate job performance has been explored by \cite{schaule2018employing} who demonstrate that wearables can be used to detect when a person is focused on her work via physiological features.
Another approach for estimation of job performance is based on other personality and individual traits \cite{barrick1991big,eysenck2012model,salgado1997five,tett1991personality} with varying degrees of success. This includes the use of conscientiousness \cite{higgins2007prefrontal,barrick2001personality,tett1991personality}, extraversion \cite{eysenck2012model}, and others.
\cite{schmidt2014general} also found a link between personality, cognitive ability, and other traits on job performance.
However, when it comes to estimation of job performance it is more common to rely on various types of questionnaires including self-reports, and supervisory and peer evaluations \cite{barrick1991big,campbell1990modeling,sonnentag2008job}.  
Job performance varies depending on demographic information (e.g., age, gender) \cite{ng2008relationship} and individual traits (personality, emotional intelligence) \cite{chiaburu2011five,kamdar2007joint}. 
It has been widely reported that anxiety is affected by context, e.g., \cite{dalal2014within,spielberger2009assessment,Spielberger:83}.
Cognitive ability, personality, affect, and anxiety are all not only related to each other but can both affect and be affected by job performance, and physical and health variables.

Mobile and wearable sensor data are powerful sources of information about human behaviour which could help us identify patterns of daily activities and human mobility \cite{madan2010social,yuan2013we,calabrese2013understanding}, but also wellness \cite{Mariani:2013}, and job and academic performance \cite{schaule2018employing}. 
Machine learning models can achieve high accuracy on very specific tasks on very small samples, e.g. estimate work load category using wearable on a cohort of 20 academic participants \cite{schaule2018employing}.
Individual perceptions of job performance are, however, hard to predict using wearable data even in small samples and specific work locations and environments \cite{olguin2009capturing}, as opposed to various work locations which is our problem at hand.
Thus, it is important to create models that provide us with the information to augment human capabilities by providing individuals with both self-monitoring and assisting technology and with the tools to better understand team work. Our paper provides a model for a global outlook of an individual well-being from physiological, psychological, and work related performance points of view. To our knowledge ours is the first model that jointly predicts  health, job-performance, and psychometric information and wellness of individuals.

Finally, the present work is a comprehensive and personalized analysis of health, psychometric information, and job performance instruments from the longitudinal Tesserae  Project\cite{mattingly2019tesserae}. An initial analysis based solely on job performance was presented in \cite{mirjafari2019differentiating}. This work reported a model to differentiate low from high performance but did not provide an estimation of the actual job performance instrument value. Also, \cite{mirjafari2019differentiating} reported predictions of daily instruments administered periodically to participants. On the contrary, our analysis is done on the single initial battery of \numberOfStandardizedTests~standardized tests used to derive the ground truth variables, not only of job performance, but of all other instruments as well.

\section{Data Collection and Description}
\label{sec:data-collection}


From Fall 2017 to Summer 2018 we recruited \numParticipants~   individuals working in knowledge 
fields in the US as part of a large-scale longitudinal research study. We collected data from these participants for a period of one year starting from January 2018. Our Tesserae project was conducted in accordance with the Institutional Review Board 
and similar authorities of all the institutions involved. Thus, we ensure the protection of the rights and welfare of human research subjects. 
No Personal Identifiable Information (PII) was shared.

\begin{table}[]
\centering
\begin{tabular}{c|r}
Cohort & \# Participants \\
\hline
1      & 217             \\
2      & 138             \\
3      & 21              \\
4      & 147             \\
5      & 31             
\end{tabular}
\caption{Participants per Cohort Used for Modeling}
\label{tab:cohorts}
\end{table}

It is important to highlight the challenge of doing predictions on this dataset because of its heterogeneity. A subset of participants was selected for external validation and was not considered during creation of our model. The remaining 554 participants came from various organizations in the USA and can be grouped in five cohorts, as shown in Table \ref{tab:cohorts}.  
A group of 217 participants work for a multinational consultancy company, a group of 138 participants work for a multinational technology company, a group of 21 participants work for a local software company, a group of 147 participants work for various smaller companies, and a group of 31 participants work for a local university. 

Another source of heterogeneity, particularly at the job-performance level, comes from the participants' roles. 254 and 297 participants self-described holding a supervisory and non-supervisory role, respectively, and 3 participants declined to mention their role within their companies. The individuals' participation was optional but those that opted in received a monetary incentive to stay in the study and comply with the data-collection protocols. This monetary compensation varied according to the compliance levels and was allocated throughout the year of study.  The monetary compensation for participants was also specific for one of the companies.  

The data collection protocols could be classified into two stages. 
An initial set of surveys used to collect the initial battery of ground truth variables, and predictors. 
A daily data-gathering that itself consisted in 
the collection of data streams from various sensors and systems on the Web (daily varying predictors). 
In the present work we are interested in the analysis of the initial ground truth battery using both the initial 
predictors as well as the daily sensor data streams. 

\subsection{Data Sources - Sensing Streams}
\label{sec:data-sources}

In order to model individuals' behaviors and physical attributes, we selected several modalities that collect in an unobtrusive way the
physiological, psychological, behavioral, and physical states of individuals; their offline
and online interactions; their phone, and social media activity and
workplace routines; and health and well-being both at work and at home. Specifically, we used a wearable to capture an individual physical, physiological, and health state. In order to capture the context of an individual's actions we use a phone agent (app) and beacons that allow as to identify the individuals' locations (home/work) during the day and mobility patterns. Finally, in order to capture higher level psychological information we use social media posts that, together with the wearable and phone agent data, could provide insights about a person's psychological states. All information is anonymized to protect the participants privacy.
We used the features extracted from these modalities as predictors.
In addition to features extracted directly from the sensors,
we also considered as predictors to features derived, composed or transformed signals obtained from the four sources. 

\emph{Wearable: Garmin Vivosmart 3}. The Garmin vivosmart 3 wristband \cite{Garmin} is a popular smart wristband (a wearable gadget) that is widely used as a fitness, activity, and wellness monitoring device by people all over the world. The type of information that this device collects include both physical/physiological (such as heartrate, step count, number of floors climbed, calories burned, physical activity such as running, walking, etc. sleep quality - including duration of light, deep, REM sleep, and total sleep periods) and psychological (such as stress -- which is based on physical signals, i.e. heart rate) \cite{Garmin}. The wearable needs to be paired via bluetooth with Connect, a Garmin App that participants install in their phones. It is also paired with an App we developed for our study (see PhoneAgent below). Both apps collect data from the wearable which is collected from our data-stream-collecting servers and into databases that anonymize and encrypt the data. We compute daily summaries from each of the signals collected from the wearable. 

 \emph{App: PhoneAgent}. As mentioned above we created an app (the PhoneAgent) for both iOS and Android devices. The app runs in the background and periodically collects 
 data saving it temporarily as JSON files that are transmitted to servers when the phone is connected to WiFi. The data collected by the PhoneAgent includes location, physical activity (walking, bicycling, driving, etc.), location, phone usage (e.g., lock/unlock) and ambient light levels. The PhoneAgent app also connects to both the wearable and the beacons (described below) via bluetooth. From the wearable, the app collects more fine-grained and real-time data than the collected by Garmin's Connect app, which includes the following data: heart rate (HR) time series, steps, floors climbed,  calories burned, and stress levels time-series. From the beacons, the app collects information about the proximity of an individual (through a key-chain beacon and backpack beacon) to either of the fixed beacons (home, office). It also provides details of interactions, such as the strength of the signal as described next. 
 
\emph{Beacons: Gimbals}. 
Beacons are low energy devices that transmit and receive Bluetooth signals to and from other devices \cite{Gimbal}. 
We use four Gimbal beacons per participant in our study. Two beacons are static Gimbal beacons \cite{Gimbal} that are placed, one at the participant's home (bedroom) and another one at her office.
The other two beacons are small, coin-size, mobile beacons that participants carry, one in their key-chain or wallet, and one in their backpacks. Beacon signals are detected by the phone through our PhoneAgent app which uses a Gimbal API library to detect proximity to the beacons. 
When a PhoneAgent enabled smartphone approaches a beacon, the phone will detect a Bluetooth signal and will collect the signal strength which is inversely proportional to the distance between the phone and the beacon. Thanks to the personal mobile beacons, this provides information about the location of an individual to their home or work beacon or to other participants. This allows us to derive features that describe the mobility of the individuals as well as other daily routines. All these features are stored by the PhoneAgent into a server and a copy of the interactions is also saved on Gimbal servers.

\emph{Social Media}: 
During the recruiting process we requested participants' access to their 
accounts on Facebook and LinkedIn, which was mandatory if they had accounts. 
As in all of the other data sources, we respect the privacy of the participants and not only anonymized their data but modified the representation so as to avoid storing raw information that may affect their privacy. 
For the present analysis we considered 
5,075 raw features computed from the Facebook data of the participants. However, a dimensionality reduction step was applied to select only the relevant features, as detailed in Section \ref{sec:feature-ranking-selection}. These raw features corresponded to a variety of categories --- 1) psycholinguistic attributes \cite{tausczik2010psychological} (that captured their language usage across keywords related to affect, cognitive attributes, perception, interpersonal focus, temporal references, biological concerns, and social and personal concerns.), 2) open vocabulary n-grams (the 5,000 most frequent uni-, bi-, and tri-grams used by the participants), 3) sentiment in posts, and 4) social capital (by measuring the activity participants' conduct and the engagement that they receive on their social media posts, for example check-ins to places, posting and sharing updates, uploading media, changing relationship status, hanging out with friends, etc.).

\subsection{Predictors}

We use a total of \numIndepVars~
candidate features (filtered out later with dimensionality reduction and feature selection techniques -- as detailed below) based on the sensor data from the PhoneAgent, Garmin wearable, Gimbal beacons, and social media. We use the wearable to extract additional information originated from the two time-series per participant: the heart rate, and stress measurements. We used these time series as separate components to extract features that facilitate discriminatory prediction based on signatures extracted using a higher order network (HON) approach, one HON per time series. We also used the heart rate to build an additional component for the ensemble using a special representation for the patterns in that time series.
\begin{table}[]
\begin{tabular}{llr}
\hline
Source                             & Sub-Modality                                    & \#             
\\ \hline

\multirow{4}{*}{Wearable}          
&  Higher Order Network - Heart Rate                         & 5$^*$                 \\ \cline{2-3}
&  Higher Order Network - Stress                        & 5$^*$                 \\ \cline{2-3}
&  Heart Rate                         & 28                 \\ \cline{2-3}
&  Other Physical                         & 26                 \\

                                 \hline
                                 
\multirow{5}{*}{Phone App}          
&  Physical Activity                         & 19                 \\ \cline{2-3} &  Context                         & 8                 \\ \cline{2-3} &  User State                         & 47                 \\ \cline{2-3} &  Phone Usage                         & 56                 \\ \cline{2-3} &  Regularity                         & 580                 \\                                 
            
\hline
\multirow{3}{*}{Beacon}          & Work Activities                       & 16                 \\ \cline{2-3} 
                                 & Other                                   &  7                  \\ \cline{2-3} 
                                 & Home Activities                               &  5         \\ \hline
\multirow{1}{*}{Social Media}
                                 &                         &  200$^*$              
         \\ \hline
\end{tabular}
    \caption{Low-level Sensor-Derived Features.}
    \label{tab:general-features}
\begin{small}$^*=$post PCA\end{small}
\end{table}
Table \ref{tab:general-features} details the number of features used per data source. 
For all features collected as time series we compute the daily mean, median, mode, minimum, maximum.

Examples of features  collected from Garmin (through the Connect API) include stress, sleep (duration for light, deep, REM sleep) and bed time, daily step counts, daily floors climbed, physical activity (duration of light, medium, heavy activity), calories burned, stress level (in range 0-100).

Example of features collected by PhoneAgent include phone usage (number of locks and unlocks, duration of locks and unlocks, etc.) daily aggregations of physical activity  such as mobility features (places visited, distance traveled, duration of sedentary state, driving, biking time, etc.)
The PhoneAgent also collects fine-grained data from the wearable such as: heart rate, sleep, stress and steps. 
In each case of  time series features we partitioned them at a daily level (which we call \emph{epoch-0}) but also in {epochs} within the day: early morning (12am - 9am), day (9am - 6pm) and evening (6pm - 12am). The objective of this partition is to identify differences of behavior during the day for the times that are associated with sleep, work, and night activities.

Example of features collected through the beacons include various measurements of closeness of the static and mobile beacons. These features in their raw form do not provide direct insights about the participants' activities but in combination with the type of beacon and the duration of the beacon interactions we capture indoor daily information such as the time spent at work (total duration a participant spends at work from the first to the last sighting of the work-beacon), the time spent at desk (percentage of the time a participant spends at their desk), the number of breaks taken away from the desk that exceed periods of 5, 15 and 30 minutes (captured by gaps in work-beacon sightings).

In our study we experimented with various time resolutions to derive the summary statistics as the distributions may have non-linear relations that may not fully capture the individuals behavior. We report  predictions for individuals with at least 2 weeks of data. 
Finally, we construct higher order network representations of people's behaviors through the heart rate and stress time series as we describe in Section \ref{sec:model}.

The predictors are highly heterogeneous due to the multi-modal nature of our dataset. This made it necessary to apply ensemble-learning strategies. Furthermore, the heterogeneity of noise in the features was not only due to the multi-modality but also due to the compliance of the participants, the quality of the data transfer, and the missing data.

\subsection{Missing Data}

In addition to the heterogeneity of the data sources, the main challenge of doing prediction with our data set was due to missing values. The data sources most affected by \emph{feature missingness}, i.e., missing values of specific predictors, were the wearable and the PhoneAgent. In particular, missingness in the later was critical as the PhoneAgent was used to collect data from the wearable and the beacons.

\emph{PhoneAgent}. Missing data in this case was mostly due to technical issues. In particular, 
keeping the phone agent running is difficult across the variety of phone models and operating system versions. Also, some adjustments were needed on the PhoneAgent because both the Garmin platform and the beacons were not recording data properly in some iOS versions.

\emph{Wearable.} Missingness was due mostly to breakages (strap, screen), lost chargers, lost wearable itself, problems with the wearable itself (e.g., did not hold charge, did not charge at all, data did not sync, unusual report of floors climbed, inability to connect to the phone), and one participant reported an allergic reaction to the nickel in the buckle.

\emph{Social Media}. It presented two challenges: not all individuals had Facebook accounts, and level of engagement of individuals in their online profiles was varied.

\emph{Beacons}. Their main challenge was the wrongly placed devices, e.g., home and office beacons were swapped and some individuals worked from home. Thus, extracting meaningful features related to location was challenging.

Finally, in addition to feature missingness, a major challenge is \emph{full-modality missingness}, i.e., participants with information missing for the entire modality, as in the case of social media, were no data was available for most participants. In such cases we resorted to group imputation tactics as detailed next.

\section{Joint Prediction Model}
\label{sec:model}

\begin{figure}[t]
	\centering
	\includegraphics[width=3in]{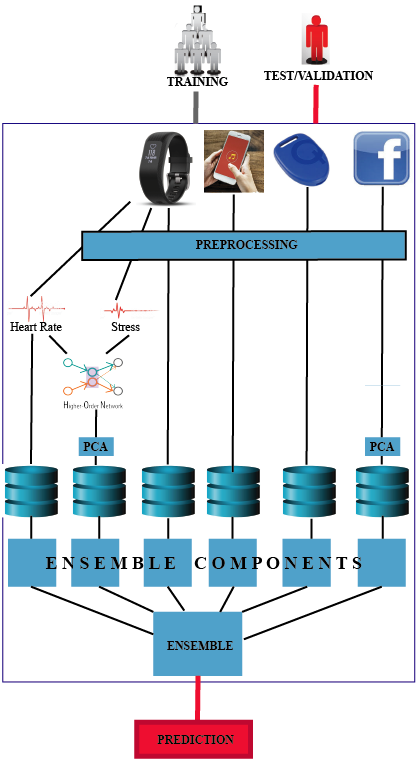}
	\caption[Ensemble Model]
	{Model Diagram: We use a set of weak modality-based learners to produce a strong prediction for each variable}
	\label{fig:sensors}
	\vspace{-4mm}
\end{figure}

We developed an ensemble learning method for joint prediction of the physical, psychological, and job-performance variables. 
As we detailed in Section \ref{sec:data-sources}, the data sources include: social media, Garmin wearable, phone agent, beacon data. Additionally, we computed heart-rate variability and used it as a separate stream and we constructed a HON based on heart rate and wearable/sensor stress measures as detailed below. Within each model, a set of candidate models are trained per ground truth variable as outlined in our model's schema in Figure \ref{fig:sensors}. The components of the ensemble are supervised techniques (regression and classification) as detailed next.
In order to deal with the complexity of the data as well as the missingness we consider four elements of our model: a) the ensmeble components that identify both linear and non-linear partitions and regressions, b) the pre- and post-processing that ensure generality and avoid outliers, c) the feature selection that eliminate redundant dimensions and selects relevant features, d) a higher-order representation of temporal data that extracts non-Markovian patterns (long-term temporal dependencies), e) imputations, both at the feature and modality level, f) a fusion strategy for the various modalities, and g) an algorithm to coordinate all these strategies as a model selection framework.
\subsection{Design of Components}
\label{subsec:components}
Our goal in designing the components was to automate the discovery of various types of variable relations and data separability for both linear and non-linear cases: linear, multicolinear, nonlinear relations. In each case we consider low and high dimensional cases. 
Thus, we considered the following regression methods as candidates for the components: linear regression (low-dimensional cases), linear regression with $L_2$-norm (multi-colinearities cases), linear regression with built-in cross validation with $L_2$-norm (high dimensional multi-colinear relations), lasso model with least angle regression (high-dimensional linear cases), Bayesian ridge regression (high dimensional cases), support vector regressor (SVR) with either linear, radial basis function, or polynomial kernel (for linear and non-linear high-dimensional relations). Finally, decision trees (CART), and random forest regression were used for non-linear relations. The selection of the optimal technique and corresponding features was done using cross-validation, as detailed below, which allows to pick the best performer per ground truth variable. The best performers were then used for training and prediction.
Likewise, we used classification counter-parts for linear and non-linear separability and high vs. low-dimensional problems. Specifically,
 we considered classification approaches for prediction including: nearest-neighbors, linear support vector machine, support vector machine with radial-basis function, decision trees, and random forest. 
At a lower level, the feature extraction was done through various intermediate steps: transformation, mapping, dimensionality reduction, fusion of sub-datasets, and feature selection.
The operations performed to build the components of the ensemble are detailed in the next sections.

\subsection{Pre- and post-processing}
\label{sec:additional-details}

\subsubsection{Cross Validation}
To guarantee generality of our models we 
used 5-fold cross validation for both model design and for final predictions. We considered both static and dynamic partitioning of the data. In order to maintain a homogeneous experimental setting we considered a fixed partitioning so that the models created across the experiments in the various experimental stages could be comparable. Thus, this static partition was applied across all the variables. However, we also considered dynamic partitioning when evaluating new models, feature selection algorithms, and techniques before the final comparison of performance which was done on the statically partitioned sets.
Data imputations were also done using the static 5-fold partitioning by imputing data in a per-fold approach. This was done to avoid overfitting due to data-leakage at this level.

\subsubsection{Outliers}
We perform an outlier analysis where out of range errors for sensing streams were analyzed for extraction issues (the most common case) resulting in integrity checks and script validation from the raw sensing streams. Examples include measurements such as sleep time on the order of 1k+ hours, negative commute times, and various other aspects. Error sources included typos in the enrollment / ingestion process, re-assigning of devices from dropped participants to newly enrolled participants, and several race / edge conditions with respect to the enrollment / ingestion process. 

\subsubsection{Data Range Transformation}
We applied systematic verification to ensure the predicted values are not outside of the prescribed ground truth ranges. Code corrections have then been applied to properly bound results as well as exploring the root cause for such out of range results.

\subsection{Feature Selection}
\label{sec:feature-ranking-selection}

A sequential exploration of various combinations of features to identify a set of predictive features per construct was conducted. The result was a curated subset of features. First, we introduce the social media feature selection process. It is worth noticing that raw social media data is not shared nor processed for privacy purposes. We only used reformatted features to remove personally identifiable information (PII).

The relevant social media data features were selected using principal component analysis (PCA) to identify the top 200 (latent) features for predicting the ground-truth variables -- the rationale was to capture complex behaviors latent in the data, and which are not directly observable in the raw signals. These 200 PCA components were derived from a total of 5,075 raw features computed from the Facebook data of the participants. These raw features corresponded to a variety of categories --- 
psycholinguistic, n-grams, sentiment posts, and social capital, as detailed before. 

The features from other sources were treated under the same selection policy to define the set of models (components). This involved five stages of selection, in addition to the feature pre-selection 
and social media selection.
First, features were selected based on correlations per fold during cross-validation. Second, features were selected by the individual candidate models. Third, a selection was done on the overall final training by the best model. Fourth, a subset of latent features was mapped using PCA for specific feature sets. Lastly, we ranked the models on predictive performance and chose the best model as our final model for each daily construct.

\subsection{Higher Order Networks (HON) of Temporal Data}

Most real sequential data does not fulfill the Markov property \cite{Xue1600028}. HON's are powerful tools that allow us to overcome this challenge by representing high order dependencies. Non-Markovian patterns in date provide unique information about the problem under study. For this reason we use a HON algorithm to provide a multi-scale representation of sequential data on a per-feature basis (e.g. heart rate and sensor-measured stress). When extracting features in sequential data, conventional methods (e.g. Markov model) might lead to information loss on the state transition with the assumption that the next status only depends on the current status. To address this limitation, we utilized a HON method to make a sufficient representation by exploring higher order dependencies in sequential data. Building the HON model consists in the following steps. 

\begin{figure}[t]
	\centering
	\includegraphics[width=3.5in]{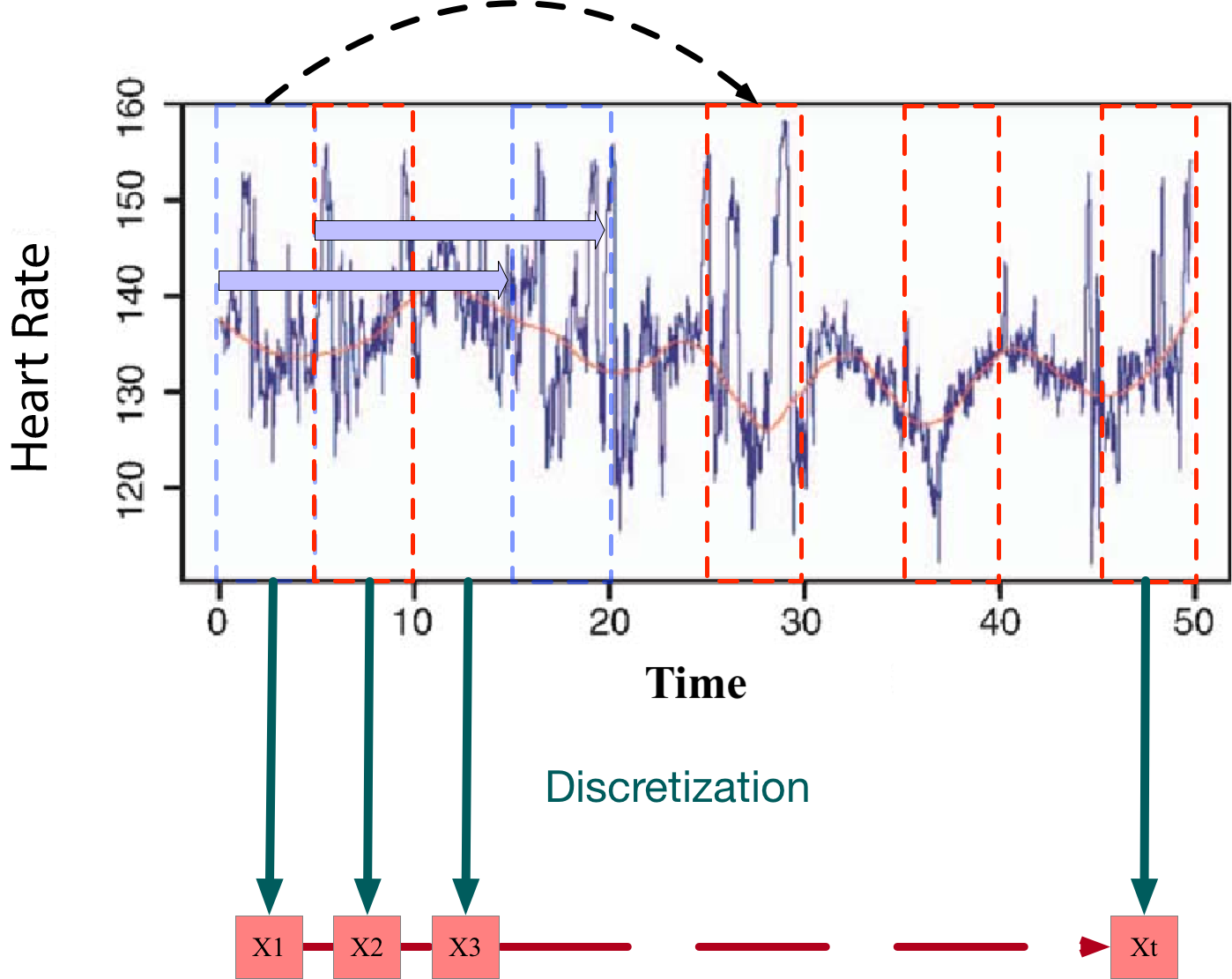}\\
	\caption[HON Creation]
	{Discretization, Satabilization, Regularization}
	\label{fig:hon-discretization}
\end{figure}

First, we apply discretization to the time series. 
The usefulness of this approach is illustrated in Figure \ref{fig:hon-discretization}. The discretization step works as a pattern recognition technique that identifies regularities in the time series that are grouped to remove high frequency components.
Since the network representation of the time series (e.g. heart rate) is not directly available, we first discretized the raw data to construct a network as shown in Figure \ref{fig:hon-discretization}. We divided time into equal- size (half hour) time slots. $x_i$ is the state in i-th time slot, which denotes the mean value for the heart rate during the corresponding time slot.

Given the discretized heart rate data, the output is the conditional probabilities of each individual $$P(x_t|x_{t-n},\ldots,x_{t-1})=\frac{\mathrm{I}(x_{t-n},\ldots,x_{t-1},x_t)}{\mathrm I(x_{t-n},\ldots,x_{t-1})}$$
where n denotes the network order, $\mathrm I(x)$ indicates the number of occurrences of $x$.

\begin{figure}[t]
	\centering
	\includegraphics[width=2in]{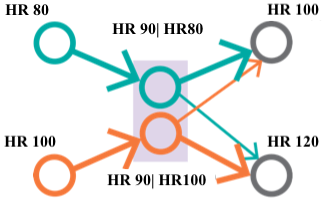}\\
	\centering
	\caption[HON ]
	{HON - Heart rate case example}
	
	\label{fig:hon}
\end{figure}

HON applies a low-pass filter to ternary relations among the selected patterns derived from the discretization step in Fig. \ref{fig:hon-discretization}. Take, for instance, the heart rate time-series illustrated in \ref{fig:hon}. An algorithm that only identifies first order relations could describe the probabilities of going from a heart rate of 80 bpm to 100 vs 120. On the other hand, HONs can differentiate patterns of heart rate transitions that go from 80 to 120 
if the previous heart rate was 80 bpm as different than if the previous heart rate was 100 bpm. PCA was used to reduce feature dimensions to a target $n\_{component} = 5$ from $727$ original features (transition probabilities).

Finally, we investigated different small orders (1-5) of HONs. The number of transition probabilities exponentially increases as the order of the network increases, which lead to a sparsity problem. In particular, most transition probabilities of each individual might be zero as the order increases. This is because, as the number of elements in a possible transition increase, the transition may not be associated to the participant.

\subsection{Imputation}

For the purpose of data imputation, two approaches were used: (1) a theoretically-driven approach that attempted to fuse data across multiple sensing streams using the knowledge of subject-matter experts (ex. sleep can be fused between the wearable, smartphone, etc.) and (2) a data-driven approach that can vary across the various features (impute via the mean, impute via zeros, etc.).
For the joint prediction of the physical, psychological, and job-performance variables, we also performed sensor-wide imputation. For this purpose, we considered the data from one stream and performed clustering on it. This allowed us to impute missing data in one stream from data in another based on the relationships between sensor streams. Other techniques applied include mean and median value imputation. 
We also performed data imputation using individual rolling means, i.e., individual mean value up to the specific moment. If there was no record at all, we filled in values using the global mean.

The level of sparsity was critical for the phone agent data at the raw data level. However, this was overcome by carefully selecting regularity-based features. Regularity features capture rhythms and routines of the various behaviors of a participant, namely the patterns within hourly phone usage, physical activity and mobility across the participant's time series. Additionally, we had to deal with sparsity for heart rate variability (HRV) when the size of the window used to compute the HRV was not adequate. Some sparsity was also due to data quality issues. Since HRV windows are calculated using Beat-to-Beat-Interval(BBI), many windows did not have a minimal number of BBI readings. This was due to inconsistencies in the data updates. HON selection also had sparsity constraints, as higher order networks provided no further information than lower order ones. 
Finally, this version of our software includes fusion of sensor data. 
Namely, we combined the features from each stream/data source and then we applied our regression models for prediction purposes.

\subsection{Fusion}

For the joint prediction of the physical, psychological, and job-performance variables, we also used a feature fusion method to combine the various modalities. The features from each streams/data sources are combined and fed into our regression and classification models.
The idea is to obtain not one but several moments from the distribution of features that provide a summary of each of the modalities. For the case of features that were numerical we use summary statistics: mean, median, standard deviation, minimum, maximum of the distributions. For the case of time series data we use the features extracted from HON, and from other summary statistics. For the phone-agent we considered regularity high-level representations, as well as the imputed values that help model building at the component-of-the-ensemble level. The specific prediction models as well as the relevant features were selected by the cross validation process. For the final ensemble we consider a model selection approach as follows.

\subsection{Model Selection}

Using the elements described so far, we build the components of the ensemble learning model by combining the HON features (heart and stress), heart rate, social media, beacons, phone agent, and wearable. We do so with the following steps:

\begin{enumerate}
    \item \emph{Feature pre-selection.} We use both the sequential exploration of various combinations of features to identify a set of predictive features per construct and the social media anonymization of features described before.
    \item \emph{Relevance-based feature selection.} Each specific technique uses an a-priori relevance (measured by correlation) on training set (either linear or non-linear correlation).
    \item \emph{Model selection.} Automated machine learning methods are applied to decide the best set of features along with the best classifier/regressor per construct
    \item \emph{Proxy ground truth.} We considered the predicted values for audit and OCB in order to perform prediction of other values. We then use these predictions and loop back to the previous step.
\end{enumerate}

Additionally, dimensionality reduction trough principal component analysis (PCA) was applied on HON construction (both stress and heart rate measures from the wearable) and on social media data. The candidate-components were described in Section \ref{subsec:components}. Thus, the main training, test is done as shown in Algorithm \ref{alg:alg1}.

 \begin{algorithm}
 \caption{Joint Model}
 \label{alg:alg1}
 \begin{algorithmic}[1]
 \renewcommand{\algorithmicrequire}{\textbf{Input:}}
 \renewcommand{\algorithmicensure}{\textbf{Output:}}
 \REQUIRE Multimodal Data $D$
 \ENSURE  Predictions
  \STATE Divide $D$ in training and validation sets $T,V$
  \STATE Use $T$ 
  to apply feature selection (top $20$ features per modality with highest correlation to the \numDepVars~ground truth variables) to select candidate features
  \FOR {\textbf{each} ground truth variable}
  \FOR {\textbf{each} \emph{parameter-set}}
  \FOR {fold $=1$ to $5$ of $T$}
  \FOR {\textbf{each} \emph{candidate-component}}
  \STATE Predict on current fold using candidate-component trained on the remaining folds
  \ENDFOR
  \ENDFOR
  \STATE \emph{SelectedComponent} $\leftarrow$ candidate with highest score across the folds
  \STATE Add \emph{SelectedComponent} to \textit{Ensemble}
  \STATE Add the fold-wise \emph{SelectedComponent} predictions $F$
  \ENDFOR
  \ENDFOR
  \STATE \emph{model} $\leftarrow$ train the ensemble on $T$
  \STATE Set predictions $P\leftarrow$ Predict($V$,\emph{model})
 \RETURN $F,P$ 
 \end{algorithmic} 
 \end{algorithm}

\section{Experiments}

We evaluate our approach using four sets of experiments. First, we investigate the performance of our model when compared to a baseline constructed with estimators derived from the ground truth values as detailed below. Second, we verify the bivariate criterion validity of the estimates of one subset of the ground truth variables using another subset. For this we consider ground truth features that are known to be predictors of job performance and compare the prediction using their sensor derived counterparts. Third, we verify the model reliability under 5-fold cross validation to ensure the models selected are generalizable and perform consistenly. Finally, an external team validated our results on a sub-cohort of participants whose data was totally unknown to us during model development.

\subsection{Data}

Our data comes from the initial formal battery of tests applied to the participants of our longitudinal study in the USA.
We considered the \numberOfStandardizedTests~standardized tests administered as initial ground-truth battery. These tests contain all the $19$ dependent variables for prediction. The independent variables come from various data sources assigned to each participant including: a wearable (Garmin VivoSmart 3), a phone agent (an app for iPhone and Android), 4 beacons (office, home, and two portable), and social media data (Facebook and Linkedin).

\subsection{Setup}

\subsubsection{Data Selection and Feature Set}
We perform preprocessing of all the streams previous to fusion of the features as described in Section \ref{sec:additional-details}.

\subsubsection{Metrics}

We use the Kendall's $\tau$ correlation coefficient which is a non-parametric measure of correlation based on rank statistics and, thus, assumes no specific structure of the data. Specifically, to compute $\tau$ scores we apply the General Monotone Model (GeMM) \cite{dougherty2012robust} to avoid any parametric assumptions about the 
variables under consideration.

\subsection{Results}

We applied our model to estimate the \numDepVars~variables describing the health, job performance, and psychometric and wellness attributes of the participants. The independent variables used were obtained from the sensors as described before. 
\subsubsection{Validation vs. Theory-Driven Baseline}
In our first set of experiments we verify that the performance of our model is comparable to theory-driven (survey-based) predictions. To create a baseline theoretical model we use the distributions of each variable and take the expected value of the training folds to estimate the values for the test fold. Table \ref{tab:general-results} shows the symmetric mean absolute percentage error (SMAPE) 
for each of the variables and the two models. 
As we can see there, using sensor-based estimates (our framework) leads to estimations with smaller errors when compared with a baseline that is based on surveys. This is important to highlight as our predictions are entirely based on wearable sensor and social media data and we do not use any survey or demographic data that would otherwise facilitate identification of patterns based on personal traits.
Our method shows competitive performance for job performance and psychological constructs. To verify the quality of modeling these later set of constructs, we provide evidence of the robustness of our predictions next.

\begin{table}[h!]
    \centering
    \begin{tabular}{l|r|r }
Variable  & Sensor-Based & Baseline\\
\hline
\VarIRB & 3.8  &   7.9	 \\
\VarITP & 4.6  &  9.4	 \\
\VarOCB & 6.8  &  14.2 \\
\VarDevInter & 18.7   &   32.9	 \\
\VarDevOrg & 14.8   &   28.5	 \\
\hline
\VarAbstraction &	6.4   &   13.4 \\
\VarVocabulary &	4.2   &   8.8 \\
\VarPerExtraversion &	8.3	  &   17.5 \\
\VarPerAgreeableness &	5.7   &   11.6 \\
\VarPerConscientiousness &	6.8   &  14.2 \\
\VarPerNeuroticism &	12.6   &   26.0 \\
\VarPerOpenness &	6.4   &   13.2 \\
\VarAffectPos &	6.6  &  13.5 \\
\VarAffectNeg &	11.4   &   22.2 \\
\VarAnxiety &	10.1   &   19.9 \\
\hline
\VarAlcohol &	30.6   &   70.4  \\
\VarTobacco &	92.2   &   195.6  \\
\VarPhysical &	30.8   &   68.9 \\
\VarSleep &	13.4   &   27.3 \\
    \end{tabular}
    \caption{Performance 
    SMAPE (\%) -- our framework vs. baseline}
    \label{tab:general-results}
\end{table}

\subsubsection{Job Performance - Improvement Assessment Over Participant-Oriented Baseline}

\begin{table}[t!]
    \centering
    \begin{tabular}{l|l |l}
Variable &	Theory-Based &	Sensor-Based	\\
\hline
\VarIRB &	0.277&	\textbf{0.298}\\
\VarITP &	0.287 &	\textbf{0.297} \\
\VarOCB &	0.170 &	\textbf{0.219} \\
\VarDevInter & 0.218 & \textbf{0.238} \\
\VarDevOrg & 0.332	& \textbf{0.346}	 
    \end{tabular}
    \caption{Incremental Criterion Validity. Comparison of job performance sensor based vs. theoretically-driven baseline -- expected performance: Kendall's $\tau$. Better performance is bolded}
    \label{tab:job-tau-expected-comparison}
\end{table}

In this experiment we evaluate the interplay of psychological and job-performance variables. The objective of this analysis is to identify the validity of the predictions from a psychological and sociological theoretical point of view. Thus, we compare the $\tau$ score accounted for by sensor-derived estimates, beyond what is accounted for by 
%
known predictors of job performance: personality ($\!\!$\cite{griffin2007new,anderson1998update,barrick1991big,barrick2001personality,salgado1997five,tett1991personality}) and cognitive ability \cite{schmidt1998validity,schmidt2004general}. This baseline is then compared with the estimation of job performance using our framework which is a sensor derived estimate. To asses the relevance of our estimations we considered the $\tau$-score of each job performance variable when predicted with the baseline and when predicted with the sensor-dervied estimates. Table \ref{tab:job-tau-expected-comparison} shows the expected $\tau$-score for both the theoretical baseline and our framework. As we can see there, our sensor-based model performs, in expectation,  better for all the variables. However, in order to fully validate this result we also performed a comparative analysis of the full set of predictions under a k-fold cross validation regime. Then we build the distribution of differences of estimations based on our model minus the estimations based on theory. 
The mode of the distributions of all the job performance variables lie in the region $\Delta_{\tau}>0$ where $\Delta_{\tau}$ is the difference of $\tau$ scores. This means that the performance of our model is better than the theory-based estimation for the bast majority of cases.

\subsubsection{Discriminant Validity}

We examined correlations between constructs and their predictions. The discriminant validity is shown in Table \ref{tab:discriminant-validity}. The objective was to verify whether our model sufficiently discriminated the constructs which is signaled by low inter-correlations among constructs and the predictions of other constructs (main diagonal = -). As shown in Table \ref{tab:discriminant-validity}, the model has good discriminant validity with correlations in the range $[-0.21,0.2]$. 

\subsubsection{Model Reliability}

\begin{figure}[t]
	\centering
	\includegraphics[width=3.5in]{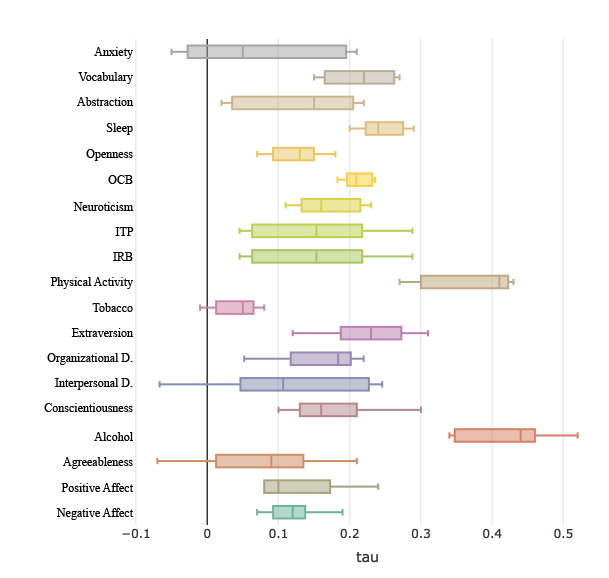}
	\caption[Stability of predictions]
	{Model Reliability: Kendall's $\tau$ confidence interval of the socio-psycho-physiological variable predictions}
	\label{fig:posterior-tau}
\end{figure}

We verified the model reliability. Figure \ref{fig:posterior-tau} shows the distribution of $\tau$ scores for each of the variables estimated. We build this posterior by running a 5-fold cross validation and thus, all the values obtained are created from independently build models. As we can see in the figure, and in agreement with previous results, the variables for which our framework performs the best are physical variables, followed by job performance and psychological constructs. Overall, the framework we propose provides consistent results across the various variables of interest, where the best performances are achieved for physical constructs and wellness being the variables. Tobacco use, and job performance are the most challenging constructs to predict. This shows our framework consistently jointly infers the health, job performance, and psychometric information and wellness of individuals.

\subsubsection{External Validation -- Totally Unknown Cohort}

We provided an external evaluation team 
with a pipeline and data to corroborate our results on a sub-cohort of participants whose data was totally unknown to us during model development. This validation was administer by MITRE Corporation. This independent evaluation lead to results consistent with what we report in Figure \ref{fig:posterior-tau}. The independent evaluation was performed on variables other than job performance and can be seen in Table \ref{tab:external}. MITRE's evaluation show
$\tau$ scores 
in the same range than the obtained in our experiments Fig. \ref{fig:posterior-tau}, is that variables such as agreeableness, neuroticism, openness, affect (particularly negative affect), and tobacco consumption are the hardest variables to predict. However, despite this challenge, a couple of these variables (openness, possitive affect), have a mean $\tau$ performance greater than $0.14$.

\begin{table}[]
\centering
\begin{tabular}{l|rrr}
Variable         & Min   & Max   & Mean  \\
\hline
\VarVocabulary       & 0.00   & 0.26  & 0.10  \\
\VarAbstraction       & 0.00   & 0.27  & 0.11  \\
\VarPerExtraversion      & 0.02  & 0.36  & 0.19  \\
\VarPerAgreeableness     & -0.09 & 0.22  & 0.07  \\
\VarPerConscientiousness & 0.05  & 0.31  & 0.15  \\
\VarPerNeuroticism       & -0.19 & 0.18  & 0.00   \\
\VarPerOpenness          & -0.11 & 0.305 & 0.14  \\
\VarAffectPos   & -0.07 & 0.32  & 0.16  \\
\VarAffectNeg   & -0.16 & 0.18  & 0.01  \\
\VarAnxiety           & 0.00   & 0.31  & 0.14  \\
\hline
\VarAlcohol             & 0.22  & 0.49  & 0.37  \\
\VarTobacco              & -0.13 & 0.00   & -0.04 \\
\VarPhysical              & 0.20   & 0.52  & 0.37  \\
\VarSleep             & 0.03  & 0.35  & 0.20  
\end{tabular}
\caption{External Validation}
\label{tab:external}
\end{table}

\section{Discussion}

In our experiments we verified the applicability of our method for jointly modeling job performance, psychometric variables, and well-being. The experiments suggest ours is a stable model with non-trivial predictive performance that is better than construct-based alternative baselines. The performance of our technique is better for physical variables of well-being such as alcohol consumption, sleep, and physical activity. The performance is also competitive for psychological and job performance variables.  
We 
verified the significance of these sensor-based predictions when compared with a participant-oriented baseline to predict job performance variables. All job variable predictions with the linear-mixed model based on our estimates produce better $\tau$-scores than the linear-mixed model created with the survey estimates. Thus our framework 
has better bivariate criterion validity.

The discriminant validity analysis show that our framework indeed sufficiently identifies the various constructs with small absolute values for the correlations [-0.21,0.2]. 
The reliability analysis shows that the model is reliable as 
a reflection of the prediction performance. Roughly speaking, 
physical variables are the most reliably sensed and estimated, followed by psychological and job performance constructs. The most challenging variables in terms of reliability are anxiety, 
tobacco consumption, 
interpersonal deviance, 
and {agreeableness}. The lower limit for the performance ranges can be explained from the difficulty of modeling social constructs in general and human performance in particular \cite{Salganik201915006}. 
These reliability results were verified externally.

It is important to highlight that our work provides a realistic assessment of the performance of prediction algorithms. This was done by respecting the nature of the data. We did not curate nor select data objects for optimal performance. Instead we worked with the original full dataset which included individuals with both full and partial set of features and modalities. Finally, the first week or two add noise (low/irregular compliance) yet our model’s 
performance 
is stable despite the missingness.
Our model provides a simpler alternative for dealing with missingness without the use of complex likelihood-based or similar techniques (e.g., \cite{mallinckrodt2003assessing,molenberghs2004analyzing}) that are otherwise necessary for predictive purposes.

\section{Conclusions}

Currently, assessing workplace performance, psychological, and physical characteristics of individuals relies either on existing full traditional questionnaires or on subjective evaluations. Furthermore, predictive techniques work on a subset of variables and work only on subsets of highly curated data and focus on only a few variables without a global overview of an individual. 
In this paper, we present the first modeling framework and benchmark that leverages sensor data from multimodal sources to jointly predict psychological, physical and physiological, and job-performance constructs. We use traditional social and psychological questionnaires to create the ground truth variables that guide the estimation of parameters of our model. We use objective mobile and personal sensing data from social media, phones, wearable and beacons as predictors and offer new insights into behavioral patterns that distinguish the various variables. We present results from an year-long study of \numParticipants~information workers collected over a period ranging from 15 days to 60 days. We created a global ensemble learning algorithm that takes advantage of various data mining techniques and feature extraction approaches to achieve this goal. 
Our results indicate that our modeling framework allows for a prediction performance above baselines. 
Despite the wealth of sources and features we use, predicting job performance and psychological constructs is a harder task than predicting physical wellness (alcohol consumption, sleep, etc.). 
Our work shows a realistic assessment of machine learning applied to 
this joint prediction task. This can also provide benefits for mitigating bias 
\cite{raghavan2020mitigating}.
Our contribution is three-fold. First, we have gathered and identified strategies for integrating highly heterogeneous data without curation, and thus, maintaining the data integrity. Second, we analyzed the different challenges in it with a systematic feature mining approach. Third, we created a benchmark for predictive tasks by leveraging the identified challenges of the real noisy or incomplete multi-modal high-dimensional data to create a comprehensive prediction and assessment of wellness: physical, psychological, and work-place well-being characteristics of individuals. Our work can be used towards the creation of more objective measures of job performance, and as a realistic and sound baseline for analysis, respectively.

\section*{Acknowledgement}

This research is based upon work supported in part by the Office of the Director of National Intelligence (ODNI), Intelligence Advanced Research Projects Activity (IARPA), via IARPA Contract No. 2017‐17042800007. The views and conclusions contained herein are those of the authors and should not be interpreted as necessarily representing the official policies, either expressed or implied, of ODNI, IARPA, or the U.S. Government. The U.S. Government is authorized to reproduce and distribute reprints for governmental purposes notwithstanding any copyright annotation therein.

\bibliographystyle{IEEEtran}
\bibliography{problesg}

\begin{landscape}
\begin{tiny}
\begin{table}[]
\centering 
  \caption{Pearson correlation among ground truth constructs (lower triangle) 
  and sensor predictions (upper triangle) 
  } 
  \label{tab:discriminant-validity} 
\begin{tabular}{@{\extracolsep{-15.50pt}}
rrrrrrrrrrrrrrrrrrrr} 
\toprule
   & \hd{irb.s}~~~~ & \hd{itp.s}~~~~ & \hd{ocb.s}~~~~ & \hd{inter.dev.s}~~~~ & \hd{org.dev.s}~~~~ & \hd{sh.abs.s}~~~~ & \hd{sh.vocab.s}~~~~ & \hd{extrav.s}~~~~ & \hd{agreeab.s}~~~~ & \hd{consc.s}~~~~ & \hd{neurot.s}~~~~ & \hd{open.s}~~~~ & \hd{pos.aff.s}~~~~ & \hd{neg.aff.s}~~~~ & \hd{stai.t.s}~~~~ & \hd{audit.s}~~~~ & \hd{gats.q.s}~~~~ & \hd{ipaq.s}~~~~ & \hd{psqi.s}~~~~ \\ 
\midrule
irb & -$\phantom{^{***}}$ & $0.05\phantom{^{***}}$ & $0.02\phantom{^{***}}$ & $0.04\phantom{^{***}}$ & $-0.03\phantom{^{***}}$ & $-0.06\phantom{^{***}}$ & $0.03\phantom{^{***}}$ & $0.03\phantom{^{***}}$ & $0.07\phantom{^{***}}$ & $0.08\phantom{^{***}}$ & $0.08\phantom{^{***}}$ & $0.01\phantom{^{***}}$ & $0\phantom{^{***}}$ & $-0.01\phantom{^{***}}$ & $-0.08^{*}\phantom{^{**}}$ & $0.06\phantom{^{***}}$ & $0.03\phantom{^{***}}$ & $0.07\phantom{^{***}}$ & $0.03\phantom{^{***}}$ \\ 
\hline
itp & $0.08\phantom{^{***}}$ & -$\phantom{^{***}}$ & $0.08^{*}\phantom{^{**}}$ & $-0.01\phantom{^{***}}$ & $-0.03\phantom{^{***}}$ & $-0.09^{*}\phantom{^{**}}$ & $-0.02\phantom{^{***}}$ & $0.11^{*}\phantom{^{**}}$ & $0.08\phantom{^{***}}$ & $0.09^{*}\phantom{^{**}}$ & $0.11^{**}\phantom{^{*}}$ & $-0.03\phantom{^{***}}$ & $0.08\phantom{^{***}}$ & $0\phantom{^{***}}$ & $-0.04\phantom{^{***}}$ & $0\phantom{^{***}}$ & $0.02\phantom{^{***}}$ & $0.11^{*}\phantom{^{**}}$ & $0.02\phantom{^{***}}$ \\ 
\hline
ocb & $-0.01\phantom{^{***}}$ & $0.2^{***}$ & -$\phantom{^{***}}$ & $0.09^{*}\phantom{^{**}}$ & $0.03\phantom{^{***}}$ & $-0.18^{***}$ & $-0.07\phantom{^{***}}$ & $0.07\phantom{^{***}}$ & $0.18^{***}$ & $-0.02\phantom{^{***}}$ & $0.11^{**}\phantom{^{*}}$ & $0\phantom{^{***}}$ & $-0.01\phantom{^{***}}$ & $0.09^{*}\phantom{^{**}}$ & $0.04\phantom{^{***}}$ & $0.01\phantom{^{***}}$ & $0.01\phantom{^{***}}$ & $0.01\phantom{^{***}}$ & $0.15^{***}$ \\ 
\hline
inter.dev & $-0.05\phantom{^{***}}$ & $-0.05\phantom{^{***}}$ & $-0.04\phantom{^{***}}$ & -$\phantom{^{***}}$ & $0.12^{**}\phantom{^{*}}$ & $0.05\phantom{^{***}}$ & $-0.02\phantom{^{***}}$ & $-0.05\phantom{^{***}}$ & $-0.02\phantom{^{***}}$ & $-0.13^{**}\phantom{^{*}}$ & $-0.01\phantom{^{***}}$ & $0\phantom{^{***}}$ & $-0.07\phantom{^{***}}$ & $0\phantom{^{***}}$ & $-0.04\phantom{^{***}}$ & $0.04\phantom{^{***}}$ & $-0.05\phantom{^{***}}$ & $-0.09^{*}\phantom{^{**}}$ & $0.02\phantom{^{***}}$ \\ 
\hline
org.dev & $0\phantom{^{***}}$ & $-0.03\phantom{^{***}}$ & $-0.02\phantom{^{***}}$ & $0.07\phantom{^{***}}$ & -$\phantom{^{***}}$ & $0.03\phantom{^{***}}$ & $0.06\phantom{^{***}}$ & $0.02\phantom{^{***}}$ & $-0.03\phantom{^{***}}$ & $-0.05\phantom{^{***}}$ & $-0.03\phantom{^{***}}$ & $0\phantom{^{***}}$ & $0.04\phantom{^{***}}$ & $0.07\phantom{^{***}}$ & $0.03\phantom{^{***}}$ & $0.09^{*}\phantom{^{**}}$ & $0.01\phantom{^{***}}$ & $0.02\phantom{^{***}}$ & $0.01\phantom{^{***}}$ \\ 
\hline
shp.abs. & $-0.06\phantom{^{***}}$ & $-0.21^{***}$ & $-0.2^{***}$ & $-0.1^{*}\phantom{^{**}}$ & $0.06\phantom{^{***}}$ & -$\phantom{^{***}}$ & $0.07\phantom{^{***}}$ & $-0.08^{*}\phantom{^{**}}$ & $-0.16^{***}$ & $-0.03\phantom{^{***}}$ & $-0.08\phantom{^{***}}$ & $-0.02\phantom{^{***}}$ & $0.01\phantom{^{***}}$ & $0.02\phantom{^{***}}$ & $-0.05\phantom{^{***}}$ & $0.02\phantom{^{***}}$ & $0.03\phantom{^{***}}$ & $-0.05\phantom{^{***}}$ & $-0.11^{**}\phantom{^{*}}$ \\ 
\hline
shp.voc. & $-0.01\phantom{^{***}}$ & $-0.09^{*}\phantom{^{**}}$ & $-0.07\phantom{^{***}}$ & $-0.13^{**}\phantom{^{*}}$ & $-0.09^{*}\phantom{^{**}}$ & $0.06\phantom{^{***}}$ & -$\phantom{^{***}}$ & $-0.08\phantom{^{***}}$ & $-0.05\phantom{^{***}}$ & $0.09^{*}\phantom{^{**}}$ & $-0.03\phantom{^{***}}$ & $-0.08^{*}\phantom{^{**}}$ & $-0.03\phantom{^{***}}$ & $-0.08\phantom{^{***}}$ & $0.01\phantom{^{***}}$ & $-0.09^{*}\phantom{^{**}}$ & $0.1^{*}\phantom{^{**}}$ & $0.03\phantom{^{***}}$ & $-0.04\phantom{^{***}}$ \\ 
\hline
extrav. & $-0.01\phantom{^{***}}$ & $0.12^{**}\phantom{^{*}}$ & $0.14^{***}$ & $0.06\phantom{^{***}}$ & $0.03\phantom{^{***}}$ & $-0.13^{**}\phantom{^{*}}$ & $-0.03\phantom{^{***}}$ & -$\phantom{^{***}}$ & $0.1^{*}\phantom{^{**}}$ & $0.02\phantom{^{***}}$ & $0.08^{*}\phantom{^{**}}$ & $-0.06\phantom{^{***}}$ & $0.13^{**}\phantom{^{*}}$ & $0.09^{*}\phantom{^{**}}$ & $-0.05\phantom{^{***}}$ & $0.17^{***}$ & $-0.02\phantom{^{***}}$ & $0.12^{**}\phantom{^{*}}$ & $-0.01\phantom{^{***}}$ \\ 
\hline
agreeab. & $0.04\phantom{^{***}}$ & $0.06\phantom{^{***}}$ & $0.04\phantom{^{***}}$ & $-0.03\phantom{^{***}}$ & $-0.09^{*}\phantom{^{**}}$ & $-0.05\phantom{^{***}}$ & $-0.02\phantom{^{***}}$ & $-0.03\phantom{^{***}}$ & -$\phantom{^{***}}$ & $0.02\phantom{^{***}}$ & $0.1^{*}\phantom{^{**}}$ & $-0.02\phantom{^{***}}$ & $-0.03\phantom{^{***}}$ & $0.01\phantom{^{***}}$ & $0.04\phantom{^{***}}$ & $-0.04\phantom{^{***}}$ & $0.08\phantom{^{***}}$ & $0.02\phantom{^{***}}$ & $0.13^{**}\phantom{^{*}}$ \\ 
\hline
consc. & $0.05\phantom{^{***}}$ & $0.03\phantom{^{***}}$ & $0.03\phantom{^{***}}$ & $-0.07\phantom{^{***}}$ & $-0.07\phantom{^{***}}$ & $-0.05\phantom{^{***}}$ & $0.09^{*}\phantom{^{**}}$ & $0.02\phantom{^{***}}$ & $0.04\phantom{^{***}}$ & -$\phantom{^{***}}$ & $-0.04\phantom{^{***}}$ & $-0.07\phantom{^{***}}$ & $0\phantom{^{***}}$ & $-0.04\phantom{^{***}}$ & $-0.05\phantom{^{***}}$ & $-0.09^{*}\phantom{^{**}}$ & $-0.03\phantom{^{***}}$ & $0.16^{***}$ & $-0.12^{**}\phantom{^{*}}$ \\ 
\hline
neur. & $0\phantom{^{***}}$ & $0.09^{*}\phantom{^{**}}$ & $0.09^{*}\phantom{^{**}}$ & $0.02\phantom{^{***}}$ & $0.02\phantom{^{***}}$ & $-0.07\phantom{^{***}}$ & $-0.06\phantom{^{***}}$ & $-0.01\phantom{^{***}}$ & $0.06\phantom{^{***}}$ & $0.05\phantom{^{***}}$ & -$\phantom{^{***}}$ & $0.04\phantom{^{***}}$ & $0.08\phantom{^{***}}$ & $0.07\phantom{^{***}}$ & $0.05\phantom{^{***}}$ & $0\phantom{^{***}}$ & $0.04\phantom{^{***}}$ & $0\phantom{^{***}}$ & $0.1^{*}\phantom{^{**}}$ \\ 
\hline
open. & $-0.02\phantom{^{***}}$ & $0.03\phantom{^{***}}$ & $0\phantom{^{***}}$ & $-0.02\phantom{^{***}}$ & $-0.02\phantom{^{***}}$ & $-0.04\phantom{^{***}}$ & $-0.03\phantom{^{***}}$ & $0.01\phantom{^{***}}$ & $0.04\phantom{^{***}}$ & $-0.09^{*}\phantom{^{**}}$ & $0.04\phantom{^{***}}$ & -$\phantom{^{***}}$ & $0\phantom{^{***}}$ & $-0.02\phantom{^{***}}$ & $0.01\phantom{^{***}}$ & $-0.02\phantom{^{***}}$ & $0.09^{*}\phantom{^{**}}$ & $-0.06\phantom{^{***}}$ & $0.14^{***}$ \\ 
\hline
pos.aff. & $0.02\phantom{^{***}}$ & $0.09^{*}\phantom{^{**}}$ & $0.09^{*}\phantom{^{**}}$ & $0.06\phantom{^{***}}$ & $-0.01\phantom{^{***}}$ & $-0.1^{*}\phantom{^{**}}$ & $-0.04\phantom{^{***}}$ & $0.13^{**}\phantom{^{*}}$ & $0.1^{*}\phantom{^{**}}$ & $0.03\phantom{^{***}}$ & $0.05\phantom{^{***}}$ & $-0.01\phantom{^{***}}$ & -$\phantom{^{***}}$ & $0.04\phantom{^{***}}$ & $-0.04\phantom{^{***}}$ & $0.11^{**}\phantom{^{*}}$ & $0.02\phantom{^{***}}$ & $0.12^{**}\phantom{^{*}}$ & $-0.04\phantom{^{***}}$ \\ 
\hline
neg.aff. & $-0.05\phantom{^{***}}$ & $0.03\phantom{^{***}}$ & $0.05\phantom{^{***}}$ & $0.04\phantom{^{***}}$ & $0.07\phantom{^{***}}$ & $-0.01\phantom{^{***}}$ & $-0.03\phantom{^{***}}$ & $0.07\phantom{^{***}}$ & $0.01\phantom{^{***}}$ & $0.02\phantom{^{***}}$ & $0.02\phantom{^{***}}$ & $0.03\phantom{^{***}}$ & $0.08\phantom{^{***}}$ & -$\phantom{^{***}}$ & $0\phantom{^{***}}$ & $0.07\phantom{^{***}}$ & $0.03\phantom{^{***}}$ & $-0.03\phantom{^{***}}$ & $0.02\phantom{^{***}}$ \\ 
\hline
stai.trait & $-0.01\phantom{^{***}}$ & $0.04\phantom{^{***}}$ & $0.06\phantom{^{***}}$ & $-0.03\phantom{^{***}}$ & $0.04\phantom{^{***}}$ & $-0.03\phantom{^{***}}$ & $-0.02\phantom{^{***}}$ & $-0.05\phantom{^{***}}$ & $0.01\phantom{^{***}}$ & $0.01\phantom{^{***}}$ & $0.11^{**}\phantom{^{*}}$ & $0.04\phantom{^{***}}$ & $0.01\phantom{^{***}}$ & $0.03\phantom{^{***}}$ & -$\phantom{^{***}}$ & $-0.01\phantom{^{***}}$ & $0.01\phantom{^{***}}$ & $-0.06\phantom{^{***}}$ & $0.12^{**}\phantom{^{*}}$ \\ 
\hline
audit & $0.07\phantom{^{***}}$ & $0\phantom{^{***}}$ & $0.04\phantom{^{***}}$ & $0.04\phantom{^{***}}$ & $0.12^{**}\phantom{^{*}}$ & $0\phantom{^{***}}$ & $0.02\phantom{^{***}}$ & $0.1^{*}\phantom{^{**}}$ & $-0.03\phantom{^{***}}$ & $0\phantom{^{***}}$ & $-0.06\phantom{^{***}}$ & $-0.08\phantom{^{***}}$ & $0.12^{**}\phantom{^{*}}$ & $0.09^{*}\phantom{^{**}}$ & $-0.03\phantom{^{***}}$ & -$\phantom{^{***}}$ & $0.01\phantom{^{***}}$ & $0.09^{*}\phantom{^{**}}$ & $-0.06\phantom{^{***}}$ \\ 
\hline
gats.q. & $-0.06\phantom{^{***}}$ & $0\phantom{^{***}}$ & $0.01\phantom{^{***}}$ & $0\phantom{^{***}}$ & $-0.07\phantom{^{***}}$ & $0\phantom{^{***}}$ & $-0.1^{*}\phantom{^{**}}$ & $-0.12^{**}\phantom{^{*}}$ & $0.01\phantom{^{***}}$ & $0.05\phantom{^{***}}$ & $0.09^{*}\phantom{^{**}}$ & $0.09^{*}\phantom{^{**}}$ & $0.02\phantom{^{***}}$ & $-0.06\phantom{^{***}}$ & $0.08\phantom{^{***}}$ & $0.01\phantom{^{***}}$ & -$\phantom{^{***}}$ & $-0.04\phantom{^{***}}$ & $0.18^{***}$ \\
\hline
ipaq & $0.06\phantom{^{***}}$ & $0.05\phantom{^{***}}$ & $0.06\phantom{^{***}}$ & $0\phantom{^{***}}$ & $-0.02\phantom{^{***}}$ & $-0.05\phantom{^{***}}$ & $0.08\phantom{^{***}}$ & $0.14^{***}$ & $0.02\phantom{^{***}}$ & $0.19^{***}$ & $0.03\phantom{^{***}}$ & $-0.03\phantom{^{***}}$ & $0.08\phantom{^{***}}$ & $0.03\phantom{^{***}}$ & $-0.04\phantom{^{***}}$ & $-0.03\phantom{^{***}}$ & $-0.02\phantom{^{***}}$ & -$\phantom{^{***}}$ & $-0.16^{***}$ \\ 
\hline
psqi & $-0.06\phantom{^{***}}$ & $0.09^{*}\phantom{^{**}}$ & $0.1^{*}\phantom{^{**}}$ & $0\phantom{^{***}}$ & $-0.02\phantom{^{***}}$ & $-0.09^{*}\phantom{^{**}}$ & $-0.09^{*}\phantom{^{**}}$ & $-0.01\phantom{^{***}}$ & $0.08\phantom{^{***}}$ & $0.01\phantom{^{***}}$ & $0.19^{***}$ & $0.1^{*}\phantom{^{**}}$ & $0.06\phantom{^{***}}$ & $0.09^{*}\phantom{^{**}}$ & $0.14^{***}$ & $0.06\phantom{^{***}}$ & $0.04\phantom{^{***}}$ & $-0.03\phantom{^{***}}$ & -$\phantom{^{***}}$ \\ 
\bottomrule
\multicolumn{11}{l}{
No * = $p > 0.05$, * = $p \leq 0.05$, ** = $p \leq 0.01$, *** = $p \leq 0.001$.}
\end{tabular} 
\end{table}

\end{tiny}
\end{landscape}

\end{document}